\documentclass[aps,preprintnumbers,nofootinbib]{revtex4}
\usepackage{amsfonts,amssymb,amsmath}            
\usepackage{graphics,graphicx,epsfig}            
\usepackage{amsthm}
\usepackage{subfigure}                              
\def\identity{\leavevmode\hbox{\small1\kern-3.8pt\normalsize1}}
\theoremstyle{plain}

\newcommand{\ket}[1]{\left | \, #1 \right\rangle}

\newcommand{\half}{\mbox{$\textstyle \frac{1}{2}$}}

\def\softt{{\leavevmode\setbox1=\hbox{t}%
\hbox to \wd1{t\kern-0.6ex{\char039}\hss}}}

\begin{document}

\title{The Impossibility of Quantum State Mirroring on Square Lattices}

\author{Alastair \surname{Kay}}
\email[]{a.s.kay@damtp.cam.ac.uk} 
\affiliation{Centre for Quantum Computation,
             DAMTP,
             Centre for Mathematical Sciences,
             University of Cambridge,
             Wilberforce Road,
             Cambridge CB3 0WA, UK}

\begin{abstract}
We discuss the possibility of engineering a Hamiltonian that can perfectly mirror a state of many excitations stored on a square lattice. In comparison to the well-studied case of a chain, there is potential for a speed-up of information transfer. However, we prove that in a large class of cases, perfect state mirroring on a square lattice is impossible.
\end{abstract}

\maketitle

\section{Introduction}

When trying to build a quantum computer, which requires us to push experimental techniques beyond their current limits, it is important to understand the trade-offs regarding how much control we require of our system. At one extreme, we have the standard formulation of a quantum computer, where one must act arbitrary single-qubit rotations on particular qubits, and perform two-qubit gates between arbitrary pairs of qubits, while still minimising the time of the interaction and the system's susceptibility to noise. At the other extreme are fixed Hamiltonian schemes, where one is assumed to have the power to engineer a system Hamiltonian whose free evolution is then capable of realising a particular quantum primitive or full-blown algorithm. In this case, no dynamical control is required (except for start and end times), and we are free to concentrate on protecting the system from decoherence, which is expected to be a lot simpler if one does not have to interact with large sections of the device.

To date, few primitives have been designed in this manner, and most research has concentrated on perfect state transfer in one dimension, but other primitives such as a long-range controlled-NOT gate \cite{Kay:2005b}, and quantum state amplification \cite{kay-2006b} (a primitive for measurement) have both been discussed, although it is also possible to implement entire algorithms \cite{kay-2007}.
The problem of quantum state transfer was first discussed by Bose \cite{Bos03}, who examined a chain of $N$ spin-\half\: particles, coupled by a nearest-neighbour exchange Hamiltonian,
$$
H=\sum_{n=1}^{N-1}(X_nX_{n+1}+Y_nY_{n+1}).
$$
The strength of the interaction between each pair of qubits was taken to be homogeneous along the length of the chain. It was numerically demonstrated that perfect state transfer was only possible for chains of 2 or 3 qubits. This was later proved rigorously \cite{Kay:2004c}. It is, of course, desirable to transfer states over greater distances. It has subsequently been demonstrated that such transfer is possible perfectly \cite{Christandl,Christandl:2004a,Kay:2004c,bose:2004a,shi:2004,transfer_comment,Kay:2005e} by engineering the couplings between the qubits. Alternatively, it is possible to use un-engineered chains by sacrificing a definite arrival time, and instead having a flag to tell us when the state has arrived \cite{Bos04,Bose:2005a}. Some advantage can be gained from encoding the information to be sent \cite{haselgrove:05}. For example, Gaussian wave packets can be transmitted along chains, minimising the dispersion of the state, thereby avoiding the need to engineer the chains \cite{osborne:03}.

Although the initial aim of state transfer was just that -- the transfer of a quantum state from qubit $1$ to qubit $N$ (with the rest of the system initialised in the $\ket{0}$ state), it was quickly realised that the developed schemes achieve much more than that. The symmetry of the system actually allows transfer of a state from any qubit $n$ to its mirror, $N+1-n$ \cite{Christandl}. Furthermore, an arbitrary state across all the qubits is mirrored, modulo the application of controlled-phase gates between all possible pairs of qubits (which are an effect of the correspondence between excitations and fermions).

In this paper, we address a similar question for two-dimensional systems -- is it possible to perfectly mirror an arbitrary quantum state on a square lattice? In this case, `mirroring' does not have a unique definition, and we discuss which is most applicable. In certain cases, we are able to prove that while the transfer of a state on a single qubit may be possible, the general mirroring of a state is impossible. The geometries under which we are able to achieve perfect state mirroring have a direct effect on the times required for particular operations (for an $N$-qubit $d$-dimensional lattice, times can often be expected to scale as $\sqrt[d]{N}$). Equally, if a particular implementation of a quantum computer is based on a square geometry, and we wish to use primitives such as state transfer to enhance the computation, it might make sense to implement the primitive in the same geometry.

\section{State Mirroring}

Paralleling the case of one-dimensional state transfer, we shall discuss Hamiltonians that have a tunable exchange coupling between nearest neighbours on an $N\times N$ square lattice,
$$
H=\sum_{i=1}^{N-1}\sum_{j=1}^{N}J_{i,j}(X_{i,j}X_{i+1,j}+Y_{i,j}Y_{i+1,j})+\sum_{i=1}^{N}\sum_{j=1}^{N-1}K_{i,j}(X_{i,j}X_{i,j+1}+Y_{i,j}Y_{i,j+1})
$$
The numbers $i$ and $j$ index the row and column of a particular qubit, and the strengths $J$ and $K$ are those that we engineer. Note that the exchange Hamiltonian preserves the number of excitations in the system, and a particular term simply corresponds to causing an excitation to hop from one qubit to its neighbour.

\subsection{Perfect Realisations}

\begin{figure}[!t]
\begin{center}
\includegraphics[width=0.4\textwidth]{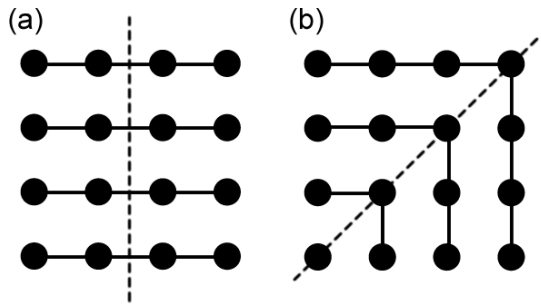}
\caption{For a square lattice of qubits (circles), state mirroring about the depicted mirror axis (dashed line) can be achieved by coupling the spins as perfect state transfer chains along the lines shown.}\label{fig:reflection}
\end{center}
\end{figure}

The possibility, or impossibility, of state mirroring in two-dimensional systems is very much dependant on the mirroring operation that we wish to achieve. For example, if all we wish to do is reflect the state in a single axis, this is a trivial construction of many parallel perfect state transfer chains, as depicted in Fig.~\ref{fig:reflection}. Hence, it only makes sense to ask about symmetries such as a rotation of $\pi$ about the centre of the lattice (which we denote by $R_{\odot}$). If we require our states to undergo the rotation $R_{\odot}$ (say), then the corresponding unitary operation is $U_\odot=e^{-iHt}$, so both the unitary and the Hamiltonian have the same eigenvectors i.e.~$H$ also has the symmetry $R_\odot$, which constrains the coupling coefficients.

In this case, a simple, and well-known, construction suffices to perfectly transfer a single excitation from any site to its symmetric point \cite{Kay:2004c,li-2005}. This is achieved by making the transfer in the two dimensions independent of each other by setting $J_{i,j}=J_{i,k}$ and $K_{i,j}=K_{k,j}$ for all $i$, $j$ and $k$, and each of the sets $\{J_{j}\}$ and $\{K_{i}\}$ correspond to a set of couplings that achieve perfect state transfer on a single chain. In essence, this works because if the excitation makes a hop in the $x$-direction (say), it still sees the same Hamiltonian in the $y$-direction, making motion in the two directions independent. However, as soon as we add a second excitation, the same argument does not hold. For example, if the two excitations were adjacent to each other, then the effective Hamiltonian applied to one of the excitations were it to step away is different to the Hamiltonian it experiences when they're adjacent -- in two dimensions, excitations can step around each other, but they can't in one.

Perfect transfer in networks of arbitrary topology has been considered in \cite{kostak}. Again, this is specifically limited to single excitation transfer. Strangely, this paper defined `nearest-neighbour' in these different topologies as the Hamiltonian being tridiagonal, which is actually imposing a one-dimensional geometry.

\subsection{The Impossibility of Mirroring}

\begin{figure}[!t]
\begin{center}
\includegraphics[width=0.2\textwidth]{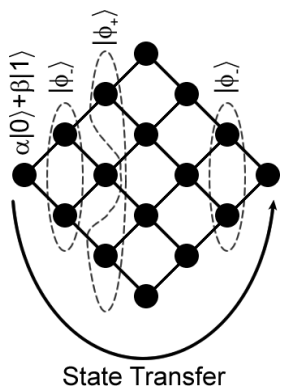}
\caption{Construction of a 0-energy eigenstate of the Hamiltonian. Arbitrary quantum states can be place on the main diagonal. States $\ket{\phi_-}$ and $\ket{\phi_+}$ are alternately placed, filling the rest of the lattice.}\label{fig:lattice}
\end{center}
\end{figure}

While we have not been able to prove that perfect state mirroring is impossible in the most general symmetry model, $R_{\odot}$, we can do so in a special case, which we denote by $R_\times$. This involves a reflection in two orthogonal axes, one of which is shown in Fig.~\ref{fig:reflection}(b). Note that this includes the uniformly-coupled case. To understand the proof of this, it is important to realise that if we act on a state $\ket{\phi_+}_{14}\ket{\phi_-}_{23}$ with the Hamiltonian
$$
(XX+YY)_{12}+(XX+YY)_{34},
$$
where $\ket{\phi_{\pm}}=(\ket{01}\pm\ket{10})/\sqrt{2}$, then we find that it is a 0-energy eigenstate. In fact, this holds independently for the $XX$ and $YY$ components, so the proof will hold for a more general class of Hamiltonians. Furthermore, $\ket{\phi_-}_{13}\ket{\chi}_2$ is a 0-energy eigenstate of
$$
(XX+YY)_{12}+(XX+YY)_{23},
$$
for any arbitrary single-qubit state $\ket{\chi}$. So, we can choose one of the main diagonals of the square lattice and place an arbitrary quantum state on it. On the nearest-neighbours to these qubits, we place (straddling the main diagonal) states $\ket{\phi_-}$. On their neighbours, we place states $\ket{\phi_+}$, and so on, until the lattice is full (as shown in Fig.~\ref{fig:lattice}). This state must be a 0-energy eigenstate of the nearest-neighbour exchange Hamiltonian with $R_\times$ symmetry. Given that it is an eigenstate, it does not evolve. The $\ket{\phi_\pm}$ components of the state are symmetric about both diagonals, so the desired initial and final states are the same. However, on the diagonal along which we have the arbitrary state, this need not be the case. For example, were we to make the initial state $\ket{100\ldots 0}$, it should evolve to $\ket{00\ldots 01}$, but no matter how we tune the couplings, this is impossible, therefore perfect state mirroring for $R_\times$ symmetry is impossible. Note that the eigenstate that we have constructed holds not only for nearest-neighbour couplings, but for any couplings over an odd distance. An equivalent construction could be made for couplings which extend over even distances, although the two cannot be combined to show that perfect mirroring is impossible for couplings over arbitrary distances.

\begin{figure}[!t]
\begin{center}
\includegraphics[width=0.2\textwidth]{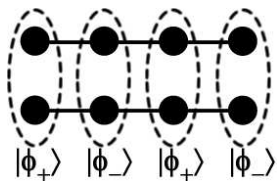}
\caption{A non-symmetric eigenstate of two identical perfect state transfer chains appears to indicate that perfect state transfer is impossible, unless we are more careful about the appearance of phase gates under exchange.}\label{fig:blocked}
\end{center}
\end{figure}

Since we have been unable to prove the impossibility of mirroring in the general case of $R_\odot$ symmetry, except for special cases such as the $2\times 2$ lattice, it is instructive to examine the reasons for this. We start by trying to generalise the approach taken so far by selecting two degenerate eigenvectors of opposite symmetry, $\ket{\psi_a}$ and $\ket{\psi_s}$. Hence, any linear combination is also an eigenvector. Let us select, for example, $\ket{\psi_T}=\beta\ket{\psi_a}-\alpha\ket{\psi_s}$. This is orthogonal to the state $\ket{x}=\alpha\ket{\psi_a}+\beta\ket{\psi_s}$, but it is not orthogonal to the mirror inverted state $\ket{\bar x}=-\alpha\ket{\psi_a}+\beta\ket{\psi_s}$, so perfect state transfer appears to be impossible. We say `appears' at this stage because one must be careful -- we have unwittingly proved that perfect state transfer is impossible in 1D as well! The fallacy comes because in the 1D case, there is a global phase on each excitation subspace which, if $\ket{\psi_a}$ and $\ket{\psi_s}$ are from different excitation subspaces, can change $\ket{x}$ into $\ket{\bar x}$. Clearly, in the 2D case, we must also allow some phases to appear. A simple example where $\ket{\psi_a}$ and $\ket{\psi_s}$ are from the same excitation subspace suffices to show that these phases should not be global on each excitation subspace -- consider two identical parallel perfect state transfer chains of even length. We would clearly like to be able to interpret their action as perfect mirroring along the central axis. However, the depicted eigenstate (Fig.~\ref{fig:blocked}) is not symmetric, and can hence be taken to be $\ket{\psi_T}$, and it appears that perfect mirroring is impossible. In this case, we can recover the action of perfect mirroring with the interpretation that some additional controlled-phase gates are applied. It seems that the only way that we can avoid this problem is to require that the $\ket{x}$ that we construct be non-symmetric with respect to the probabilities (not only the amplitudes) of finding the excitation on each site. This, in turn, imposes that the degenerate eigenvectors of opposite symmetry must come from the same excitation subspace. So far, we have been unable to find suitable pairs of eigenvectors in the general case.

\section{Conclusions}

We have discussed the possibility of perfectly mirroring an arbitrary quantum state on a square lattice using only a fixed Hamiltonian (as opposed to perfect state transfer, which concentrates on the transmission of a single excitation). In a large range of cases, we have shown that while state transfer may be possible, the perfect mirroring operation is not. This proof can be expected to hold for all even-dimensional hypercubes with similar symmetry properties. However, it remains an open question as to whether the impossibility can be proved, or counter-examples found, for the most general $R_{\odot}$ symmetry for square lattices, or for cubic lattices.

The impossibility was proven by constructing non-symmetric eigenvectors of the Hamiltonian which, since they are stationary, do not get correctly mirrored. It is worth noting that our method for constructing eigenvectors using the states $\ket{\phi_\pm}$ is applicable to a large range of alternative geometries, such as cubes and octahedra, although the eigenvectors in these cases preserve the symmetry of the Hamiltonian.


\begin{thebibliography}{17}
\expandafter\ifx\csname natexlab\endcsname\relax\def\natexlab#1{#1}\fi
\expandafter\ifx\csname bibnamefont\endcsname\relax
  \def\bibnamefont#1{#1}\fi
\expandafter\ifx\csname bibfnamefont\endcsname\relax
  \def\bibfnamefont#1{#1}\fi
\expandafter\ifx\csname citenamefont\endcsname\relax
  \def\citenamefont#1{#1}\fi
\expandafter\ifx\csname url\endcsname\relax
  \def\url#1{\texttt{#1}}\fi
\expandafter\ifx\csname urlprefix\endcsname\relax\def\urlprefix{URL }\fi
\providecommand{\bibinfo}[2]{#2}
\providecommand{\eprint}[2][]{\url{#2}}

\bibitem[{\citenamefont{Kay and Ericsson}(2005)}]{Kay:2005b}
\bibinfo{author}{\bibfnamefont{A.}~\bibnamefont{Kay}} \bibnamefont{and}
  \bibinfo{author}{\bibfnamefont{M.}~\bibnamefont{Ericsson}},
  \bibinfo{journal}{New J. Phys.} \textbf{\bibinfo{volume}{7}},
  \bibinfo{pages}{143} (\bibinfo{year}{2005}).

\bibitem[{\citenamefont{Kay}(2007)}]{kay-2006b}
\bibinfo{author}{\bibfnamefont{A.}~\bibnamefont{Kay}}, \bibinfo{journal}{Phys.
  Rev. Lett.} \textbf{\bibinfo{volume}{98}}, \bibinfo{pages}{010501}
  (\bibinfo{year}{2007}).
  
\bibitem[{\citenamefont{Kay}(2007)\citenamefont{Kay}}]{kay-2007}
\bibinfo{author}{\bibfnamefont{A.}~\bibnamefont{Kay}}
  (\bibinfo{year}{2007}),
  \bibinfo{note}{quant-ph/0702092}.

\bibitem[{\citenamefont{Bose}(2003)}]{Bos03}
\bibinfo{author}{\bibfnamefont{S.}~\bibnamefont{Bose}}, \bibinfo{journal}{Phys.
  Rev. Lett.} \textbf{\bibinfo{volume}{91}}, \bibinfo{pages}{207901}
  (\bibinfo{year}{2003}).

\bibitem[{\citenamefont{Christandl et~al.}(2005)\citenamefont{Christandl,
  Datta, Dorlas, Ekert, Kay, and A.}}]{Kay:2004c}
\bibinfo{author}{\bibfnamefont{M.}~\bibnamefont{Christandl}},
  \bibinfo{author}{\bibfnamefont{N.}~\bibnamefont{Datta}},
  \bibinfo{author}{\bibfnamefont{T.}~\bibnamefont{Dorlas}},
  \bibinfo{author}{\bibfnamefont{A.}~\bibnamefont{Ekert}},
  \bibinfo{author}{\bibfnamefont{A.}~\bibnamefont{Kay}}, \bibnamefont{and}
  \bibinfo{author}{\bibfnamefont{A.~J.} \bibnamefont{Landahl}}, \bibinfo{journal}{Phys. Rev. A}
  \textbf{\bibinfo{volume}{71}}, \bibinfo{pages}{032312}
  (\bibinfo{year}{2005}).

\bibitem[{\citenamefont{Christandl et~al.}(2004)\citenamefont{Christandl,
  Datta, Ekert, and Landahl}}]{Christandl}
\bibinfo{author}{\bibfnamefont{M.}~\bibnamefont{Christandl}},
  \bibinfo{author}{\bibfnamefont{N.}~\bibnamefont{Datta}},
  \bibinfo{author}{\bibfnamefont{A.}~\bibnamefont{Ekert}}, \bibnamefont{and}
  \bibinfo{author}{\bibfnamefont{A.~J.} \bibnamefont{Landahl}},
  \bibinfo{journal}{Phys. Rev. Lett.} \textbf{\bibinfo{volume}{92}},
  \bibinfo{pages}{187902} (\bibinfo{year}{2004}).

\bibitem[{\citenamefont{Albanese et~al.}(2004)\citenamefont{Albanese,
  Christandl, Datta, and Ekert}}]{Christandl:2004a}
\bibinfo{author}{\bibfnamefont{C.}~\bibnamefont{Albanese}},
  \bibinfo{author}{\bibfnamefont{M.}~\bibnamefont{Christandl}},
  \bibinfo{author}{\bibfnamefont{N.}~\bibnamefont{Datta}}, \bibnamefont{and}
  \bibinfo{author}{\bibfnamefont{A.}~\bibnamefont{Ekert}},
  \bibinfo{journal}{Phys. Rev. Lett} \textbf{\bibinfo{volume}{93}},
  \bibinfo{pages}{230502} (\bibinfo{year}{2004}).

\bibitem[{\citenamefont{Yung and Bose}(2005)}]{bose:2004a}
\bibinfo{author}{\bibfnamefont{M.-H.} \bibnamefont{Yung}} \bibnamefont{and}
  \bibinfo{author}{\bibfnamefont{S.}~\bibnamefont{Bose}},
  \bibinfo{journal}{Phys. Rev. A} \textbf{\bibinfo{volume}{71}},
  \bibinfo{pages}{032310} (\bibinfo{year}{2005}).

\bibitem[{\citenamefont{Shi et~al.}(2005)\citenamefont{Shi, Li, Song, and
  Sun}}]{shi:2004}
\bibinfo{author}{\bibfnamefont{T.}~\bibnamefont{Shi}},
  \bibinfo{author}{\bibfnamefont{Y.}~\bibnamefont{Li}},
  \bibinfo{author}{\bibfnamefont{Z.}~\bibnamefont{Song}}, \bibnamefont{and}
  \bibinfo{author}{\bibfnamefont{C.~P.} \bibnamefont{Sun}},
  \bibinfo{journal}{Phys. Rev. A} \textbf{\bibinfo{volume}{71}},
  \bibinfo{pages}{032309} (\bibinfo{year}{2005}).

\bibitem[{\citenamefont{Karbach and Stolze}(2005)}]{transfer_comment}
\bibinfo{author}{\bibfnamefont{P.}~\bibnamefont{Karbach}} \bibnamefont{and}
  \bibinfo{author}{\bibfnamefont{J.}~\bibnamefont{Stolze}},
  \bibinfo{journal}{Phys Rev. A} \textbf{\bibinfo{volume}{72}},
  \bibinfo{pages}{030301(R)} (\bibinfo{year}{2005}).

\bibitem[{\citenamefont{Kay}(2006)}]{Kay:2005e}
\bibinfo{author}{\bibfnamefont{A.}~\bibnamefont{Kay}}, \bibinfo{journal}{Phys.
  Rev. A} \textbf{\bibinfo{volume}{73}}, \bibinfo{pages}{032306}
  (\bibinfo{year}{2006}).

\bibitem[{\citenamefont{Burgarth and Bose}(2005{\natexlab{a}})}]{Bos04}
\bibinfo{author}{\bibfnamefont{D.}~\bibnamefont{Burgarth}} \bibnamefont{and}
  \bibinfo{author}{\bibfnamefont{S.}~\bibnamefont{Bose}},
  \bibinfo{journal}{Phys. Rev. A} \textbf{\bibinfo{volume}{71}},
  \bibinfo{pages}{052315} (\bibinfo{year}{2005}{\natexlab{a}}).

\bibitem[{\citenamefont{Burgarth and Bose}(2005{\natexlab{b}})}]{Bose:2005a}
\bibinfo{author}{\bibfnamefont{D.}~\bibnamefont{Burgarth}} \bibnamefont{and}
  \bibinfo{author}{\bibfnamefont{S.}~\bibnamefont{Bose}}, \bibinfo{journal}{New
  J. Phys.} \textbf{\bibinfo{volume}{7}}, \bibinfo{pages}{135}
  (\bibinfo{year}{2005}{\natexlab{b}}).

\bibitem[{\citenamefont{Haselgrove}(2005)}]{haselgrove:05}
\bibinfo{author}{\bibfnamefont{H.~L.} \bibnamefont{Haselgrove}},
  \bibinfo{journal}{Phys. Rev. A} \textbf{\bibinfo{volume}{72}},
  \bibinfo{pages}{062326} (\bibinfo{year}{2005}).

\bibitem[{\citenamefont{Osborne and Linden}(2004)}]{osborne:03}
\bibinfo{author}{\bibfnamefont{T.~J.} \bibnamefont{Osborne}} \bibnamefont{and}
  \bibinfo{author}{\bibfnamefont{N.}~\bibnamefont{Linden}},
  \bibinfo{journal}{Phys. Rev. A} \textbf{\bibinfo{volume}{69}},
  \bibinfo{pages}{052315} (\bibinfo{year}{2004}).

\bibitem[{\citenamefont{Li et~al.}(2005)\citenamefont{Li, Song, and
  Sun}}]{li-2005}
\bibinfo{author}{\bibfnamefont{Y.}~\bibnamefont{Li}},
  \bibinfo{author}{\bibfnamefont{Z.}~\bibnamefont{Song}}, \bibnamefont{and}
  \bibinfo{author}{\bibfnamefont{C.~P.} \bibnamefont{Sun}},
  (\bibinfo{year}{2005}),
  \bibinfo{note}{quant-ph/0504175}.

\bibitem[{\citenamefont{Kostak et~al.}(2007)\citenamefont{Kostak, Nikolopoulos, and
  Jex}}]{kostak}
\bibinfo{author}{\bibfnamefont{V.}~\bibnamefont{Ko\v s\softt\'ak}},
  \bibinfo{author}{\bibfnamefont{G.~M.}~\bibnamefont{Nikolopoulos}}, \bibnamefont{and}
  \bibinfo{author}{\bibfnamefont{I.} \bibnamefont{Jex}},
  (\bibinfo{year}{2007}),
  \bibinfo{note}{quant-ph/0702016}.
  
\end{thebibliography}

\end{document}